\documentclass[a4paper,twocolumn,10pt,accepted=2021-07-17]{quantumarticle}
\pdfoutput=1
\usepackage{graphics}
\usepackage{bbm}
\usepackage{amsmath}
\usepackage[utf8]{inputenc}
\usepackage{tikz}
\usepackage{lipsum}
\usepackage{xcolor}
\usepackage{cite}
\usepackage[bookmarks=true,bookmarksnumbered=false,bookmarksopen=false, breaklinks=false,pdfborder={0 0 0},backref=false,colorlinks=true]{hyperref}

\sffamily
\newcommand{\op}{\boldsymbol}

\begin{document}

\title{Momentum Kicks in Imperfect Which-Way Measurement}
\author{Neha Pathania}
\orcid{0000-0002-3385-4761}
\email{neha@ctp-jamia.res.in}
\author{Tabish Qureshi}
\orcid{0000-0002-8452-1078}
\email{tabish@ctp-jamia.res.in}
\affiliation{Centre for Theoretical Physics, Jamia Millia Islamia, New Delhi,
India.}


\begin{abstract}
There has been an intense debate on the question as to whether a quanton,
passing through a double-slit, experiences a 'momentum kick' due to
the act of which-way detection. There have been conflicting points of
view on this issue over many decades. This issue is addressed here in the
general setting where the which-way detection may be imperfect. 
It is shown here that the loss of interference may still be interpreted
as arising out of tiny momentum kicks which the quanton appears to
receive, irrespective of the nature of the which-way detector.
Interestingly, the magnitude of the random momentum kicks is
always $h/2d$, $d$ being the slit separation, irrespective
of how perfect or imperfect the which-way detection is. This is contrary
to what has been suggested in the earlier literature. The imperfection
of which-way detection decides how frequent are the momentum kicks. It
has been shown earlier that for perfect which-way detection, the quanton
receives a momentum kick fifty percent of the time.  Here it is shown that for
imperfect which-way detection, the quanton receives momentum kicks of the
same magnitude, but less often. A precise relation between the
frequency of kicks and the visibility of interference is found here.
\end{abstract}

\maketitle

\section{Introduction}
The concept of complementarity, introduced by Niels Bohr \cite{bohr}, is
best illustrated in the two-slit experiment with particles. Right at the time
of the formulation of Bohr's principle, Einstein proposed his famous
recoiling-slit experiment, in a bid to refute it \cite{recoil}. Although
Einstein's bid turned out to be unsuccessful, it generated a lively debate
which continues to this day. With the advancement of technology and 
sophisticated experimental
techniques, this thought experiment has now been realized in different ways
\cite{recoillui,recoildorner,utter}.  Bohr invoked Heisenberg's uncertainty
principle to give a rebuttal to Einstein by pointing
out that measuring the momentum of the recoiling slit, in order to find
which of the two slits the particle went through, would produce an uncertainty
in the
position of the recoiling slit, which in turn would wash out the interference.
Bohr's reply led many authors to assume that complementarity was grounded
in the uncertainty principle, and was probably another way of stating it.
Such belief prevailed until Scully, Englert and Walther proposed a which-way
experiment with atoms and using micromaser cavities
as a which-way detector. They claimed that the loss of interference in the
proposed experiment was purely a result of quantum correlations between the
path states of the atom and the states of cavities, and did not involve
any position-momentum uncertainty \cite{SEW}. They concluded that the
which-way detection process does not involve any momentum transfer to the
interfering particle.
Storey et.al. produced a counter argument to this claim by apparently proving
that if an interference pattern is destroyed in a which-way experiment,
a momentum of at least the magnitude $\hbar/d$ should be transferred to the
particle, where $d$ is the separation between the two slits \cite{storey}.
A momentum transfer of an amount smaller than that would not destroy
the interference completely, they argued. The focus of the debate then 
shifted to settling the question whether there is a momentum transfer to the
particle involved in the process of which-way detection
\cite{kick-englert,kick-storey,kick-wiseman1,kick-wiseman2,kick-wiseman3,wiseman2}.

\begin{figure}
\centerline{\resizebox{8.5cm}{!}{\includegraphics{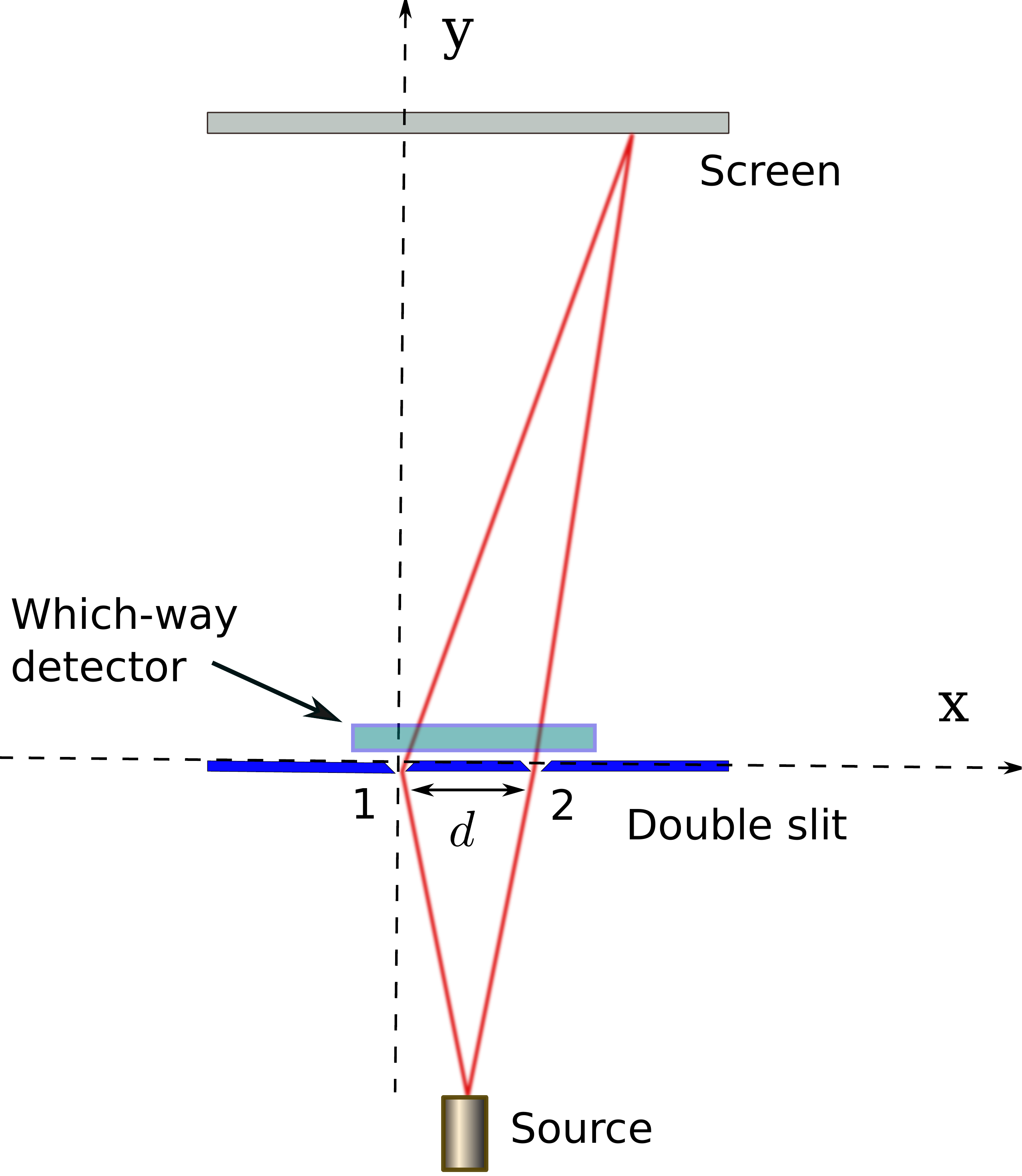}}}
\caption{Schematic diagram of a two-slit interference experiment in the
presence of a which-way detector. Slit 1 is located at $x=0$ and slit 2 is
located at $x=d$. The quanton travels in the positive $y$ direction.}
\label{twoslit}
\end{figure}

Later it was shown that the complementarity principle can be understood
in terms of the ubiquitous entanglement between the particle and the
which-way detector, and also equivalently in terms of the uncertainty between
certain operators of the which-way detector, and the well-known wave-particle
duality relation \cite{englert} can be derived from both \cite{tqrv,durr}.
However, the question whether there is a momentum transfer to the particle
or not, is still being actively studied \cite{mir,xiao}. 
Very recently it was shown that the two views which say that there is, and
there isn't a momentum kick, are completely equivalent, and depend on
which basis set of the which-way detector one is considering \cite{tq}.
However, perfect which-way detection was used in that argument. In situations
where the which-way detection may not be perfect, can one still talk
of momentum kicks as shown in the earlier result \cite{tq}? That is the
question we address in the following investigation.

\section{Imperfect which-way detection and interference}

Let us consider a quanton passing through a double-slit, as shown in 
Figure \ref{twoslit}. The traveling quanton also interacts with a which-way
detector, the detailed nature of which we do not specify. We write the
joint state of the quanton and which-way detector as the state which was
consider by Storey et. al. \cite{storey}
\begin{equation}
|\Psi\rangle = \int  dx \psi(x)|x\rangle\otimes|D_x\rangle ,
\label{ent1}
\end{equation}
where $\psi(x)$ represents the wavefunction of the quanton just as it
emerges from the double-slit, and $|D_x\rangle$ are the states of the
which-way detector. Instead of following the general analysis of Storey et. al.,
we use the fact that two slits are physically separated, and the wavefunctions
localized at each, are disjoint, and orthogonal to each other. Thus we can
write
\begin{equation}
|\Psi\rangle = \tfrac{1}{\sqrt{2}}\int dx \left[ \psi_1(x)|d_1\rangle + \psi_2(x)|d_2\rangle \right]|x\rangle,
\label{ent2}
\end{equation}
where are $\psi_1(x),\psi_2(x)$ are wavefunctions sharply localized at $x=0$
and $x=d$, respectively, and are orthogonal by virtue of their spatial
separation. We also recognize the fact the which-way detector states
$|d_1\rangle,|d_2\rangle$ may or may not have a position degree of freedom.
For example, they could just represent spin degree of freedom.
Consequently $|d_1\rangle, |d_2\rangle$ are assumed to be certain normalized
states of the which-way detector, in an effectively two-dimensional Hilbert
space. If $|d_1\rangle, |d_2\rangle$ are orthogonal, they will perform a
perfect which-way detection of the quanton. Here we assume that
$|d_1\rangle, |d_2\rangle$ are not necessarily orthogonal. For our purpose,
we omit the integral in (\ref{ent2}), and write the combined state as a
function of $x$
\begin{equation}
|\Psi(x)\rangle = \tfrac{1}{\sqrt{2}} \left[ \psi_1(x)|d_1\rangle + \psi_2(x)|d_2\rangle \right],
\label{ent3}
\end{equation}

After emerging from the double slit, the quanton travels a distance $D$ to
the screen in a time $t$, and the state at time $t$ is given by
\begin{equation}
\Psi(x,t) = \tfrac{1}{\sqrt{2}} \op{U}(t)\left[ \psi_1(x)|d_1\rangle + \psi_2(x)|d_2\rangle \right],
\label{psit}
\end{equation}
where $\op{U}(t)$ represents the Schr\"odinger time evolution operator.
A full treatment would require taking into account the y-dependence of the
wavefunction $\Psi(x,t)$. However, the evolution in the y-direction is
rather trivial, and serves only to transport the quanton from the slits
to the screen. It is independent of the evolution in the x-direction, and
thus plays no role in the dynamics of interference. For this
reason we choose to ignore the explicit y-dependence of the quanton 
wavefunction, and just assume that $\Psi(x,t)$ is the wavefunction of the
quanton at time $t$, \emph{when it has reached the screen}.
The probability density of the quanton falling on the screen at a
position $x$, is then given by 
\begin{eqnarray}
|\Psi(x,t)|^2 &=& \tfrac{1}{2} \left[ |\psi_1(t)|^2 + |\psi_2(t)|^2 \right.\nonumber\\
&&\left. + \psi_1^*(t)\psi_2(t)\langle d_1|d_2\rangle + 
\psi_2^*(t)\psi_1(t)\langle d_2|d_1\rangle\right],\nonumber\\
\end{eqnarray}
where $\psi_j(t)\equiv\psi_j(x,t)=\op{U}(t)\psi_j(x)$. 
Notable is the fact that the last two terms in the above equation, which
represent interference, are reduced by a factor $|\langle d_1|d_2\rangle|$.
In fact, it is well known that in a symmetric two slit interference
experiment, the visibility of interference turns out to be
\cite{englert,tqrv,awpd}
\begin{equation}
\mathcal{V} = |\langle d_1|d_2\rangle|.
\label{V}
\end{equation}

Now if there were no which-way detector in the path of the quanton, its
wavefunction would be given by
\begin{equation}
\Psi_0(x) = \tfrac{1}{\sqrt{2}} \left[ \psi_1(x) + \psi_2(x) \right],
\label{ent0}
\end{equation}
which would lead to maximum visibility $\mathcal{V}=1$.
It should be emphasized here that one can, in principle, choose an interaction
between the quanton and the which-way detector such that the
$\psi_1(x),\psi_2(x)$ in (\ref{ent0}) are the
same as those in (\ref{ent3}), and the which-way detection does not change
the individual states of the particle emerging from the slits.
So there is no question of any additional momentum kick or 'momentum
back-action' which has been considerably debated in the literature. This
was the point of view of Scully, Englert and Walther \cite{SEW}.

\section{Distinguishing between the two paths}

If the states $|d_1\rangle, |d_2\rangle$ in (\ref{ent3}) are orthogonal,
the two paths can be distinguished unambiguously, by simply measuring
an observable of the which-way detector for which $|d_1\rangle, |d_2\rangle$
yield different eigenvalues. However, if $|d_1\rangle, |d_2\rangle$ are
not orthogonal, no such observable exists. However, there is a method in 
which two non-orthogonal states can be distinguished unambiguously, with
the caveat that the method can occasionally fail. The success and failure
events are distinct, and one knows if one has failed or not. The result is
that one knows the events in which one succeeds, and in those, one can
distinguish between the two non-orthogonal state \emph{unambiguously}. This
procedure is called unambiguous quantum state discrimination (UQSD)
\cite{uqsd,dieks,peres,jaeger2}.

There are different ways of doing UQSD. In one method it is
assumed that the Hilbert space of the non-orthogonal states
$|d_1\rangle, |d_2\rangle$ (path-detector states in our case) is not
two-dimensional, but three dimensional, described by an orthonormal
basis of states $|q_1\rangle, |q_2\rangle, |q_3\rangle$.
The reason for choosing this method will become clear in the following analysis.
The basis is chosen in such a way that the detector states $|d_1\rangle,
|d_2\rangle$ can be represented as \cite{jaeger2}
\begin{eqnarray}
|d_1\rangle &=& \alpha|q_1\rangle + \beta|q_3\rangle \nonumber\\
|d_2\rangle &=& \gamma|q_2\rangle + \delta|q_3\rangle ,
\label{d1d2}
\end{eqnarray}
where $\alpha$ and $\gamma$ are real, and $\beta, \delta$ satisfy
\begin{eqnarray}
|\beta| |\delta| &\ge& |\langle d_1|d_2\rangle|,\nonumber\\
|\beta|^2&=& |\langle d_1|d_2\rangle| \nonumber\\
\alpha = \gamma &=& \sqrt{1-|\langle d_1|d_2\rangle|}
\label{betagamma}
\end{eqnarray}
In the expanded Hilbert space, one can now measure an operator (say)
\begin{equation}
\op{Q} = a|q_1\rangle\langle q_1| + b|q_2\rangle\langle q_2|
+ c|q_3\rangle\langle q_3|.
\end{equation}
Here $\langle q_2|d_1\rangle = 0$ and $\langle q_1|d_2\rangle = 0$,
which means that getting a measured eigenvalue $a$ of $\op{Q}$ means
that the state could not have been $|d_2\rangle$, and thus it is $|d_1\rangle$.
Similarly, getting a measured eigenvalue $b$ of $\op{Q}$ means that the
state could not have been $|d_1\rangle$, and thus it is $|d_2\rangle$.
In these two cases one can unambiguously distinguish between
$|d_1\rangle$ and $|d_2\rangle$. However, 
$\langle q_3|d_1\rangle \neq 0$ and $\langle q_3|d_2\rangle \neq 0$
in which case one cannot tell if the state was $|d_1\rangle$ or $|d_2\rangle$.
In order to make the procedure maximally efficient, one would like to
minimize the probability of getting the eigenvalue $c$, which represents
the failure of the state discrimination. The values of
$\beta,\delta$ that we have chosen in (\ref{betagamma}) are such that they
minimize the probability of failure, and \emph{maximize} the probability
of successfully distinguishing between $|d_1\rangle$ and $|d_2\rangle$
\cite{jaeger2}.

Next we substitute (\ref{d1d2}) in (\ref{ent3}) to write
\begin{eqnarray}
|\Psi(x)\rangle &=& \tfrac{1}{\sqrt{2}} \left[ \psi_1(x)(\alpha|q_1\rangle + \beta|q_3\rangle) \right. \nonumber\\
&&\left. + \psi_2(x) (\alpha|q_2\rangle + \delta|q_3\rangle) \right],\nonumber\\
&=& \tfrac{1}{\sqrt{2}} \alpha[\psi_1(x)|q_1\rangle
+ \psi_2(x)|q_2\rangle] \nonumber\\
 && + \tfrac{1}{\sqrt{2}}[\beta\psi_1(x) + \delta\psi_2(x)] |q_3\rangle.
\label{state1}
\end{eqnarray}
We would like to emphasize that by writing the state in the above form,
we have neither changed anything, nor carried out any physical process.
From (\ref{betagamma}), one can see that $\delta$ is equal to $\beta$
up to a phase factor like $e^{i\theta}$. Let us first assume $\theta$ to
be zero, so that the state (\ref{state1}) simplifies to
\begin{eqnarray}
|\Psi(x)\rangle &=& \tfrac{\alpha}{\sqrt{2}} [\psi_1(x)|q_1\rangle
+ \psi_2(x)|q_2\rangle] \nonumber\\
 && + \tfrac{\beta}{\sqrt{2}}[\psi_1(x) + \psi_2(x)] |q_3\rangle.
\label{state2}
\end{eqnarray}
We introduce two new states
$|q_{\pm}\rangle = \tfrac{1}{\sqrt{2}}(|q_1\rangle \pm |q_2\rangle)$, in
terms of which the state (\ref{state2}) can be written as
\begin{eqnarray}
|\Psi(x)\rangle &=& \tfrac{\alpha}{2}[\psi_1(x) + \psi_2(x)]|q_+\rangle\nonumber\\
&& + \tfrac{\alpha}{2}[\psi_1(x) - \psi_2(x)]|q_-\rangle\nonumber\\
 && + \tfrac{\beta}{\sqrt{2}}[\psi_1(x) + \psi_2(x)] |q_3\rangle.
\label{state3}
\end{eqnarray}
It is interesting to observe that the state of the quanton correlated to
the which-way detector state $|q_+\rangle$ and $|q_3\rangle$ is the same,
and is simply the state of the quanton that would have been, if there were
no which-way detector, i.e., $|\Psi_0(x)\rangle$ given by (\ref{ent0}).
The quanton state correlated to $|q_-\rangle$ differs from $|\Psi_0(x)\rangle$
only in that the two paths are out of phase by $\pi$.

\section{Apparent momentum kick}

When the quanton emerges from the double-slit, the state consists of
two parts each localized within the very narrow region of the slit.
Thus $\psi_1(x)$ represents a state sharply localized at $x=0$ (position
of slit 1), and $\psi_2(x)$ is sharply localized at $x=d$ (position
of slit 2). The width of these two states in position would be approximately
the same as the width of each slit. Of course, after emerging from the slits,
when $\psi_1(x)$ and $\psi_2(x)$ evolve in time, their widths will increase
very fast. At the double slit, the quanton state correlated to $|q_-\rangle$
can be written in a slightly different way:
\begin{equation}
 \tfrac{1}{\sqrt{2}}[ \psi_1(x) - \psi_2(x)] =
\tfrac{1}{\sqrt{2}} \exp(\frac{i}{\hbar}p_0x)[ \psi_1(x) + \psi_2(x)],
\label{kickstate}
\end{equation}
where $p_0 = h/2d$. In arriving at (\ref{kickstate}), we have assumed that
$\psi_1(x)$ has support only in a narrow region localized at $x=0$, and
$\psi_2(x)$ has support only in a narrow region localized at $x=d$.
The exponential function $\exp(\frac{i}{\hbar}p_0x)$ in these two regions
can be approximated by $\exp(\frac{i}{\hbar}p_0\cdot 0)=1$ and
$\exp(\frac{i}{\hbar}p_0d)=e^{i\pi}=-1$, respectively.
The factor $e^{\frac{i}{\hbar}p_0x}$ can be considered
as a momentum kick that the quanton experiences, in comparison to the
quanton state correlated to $|q_+\rangle$, which also happens to be identical
to the undisturbed state $|\Psi_0(x)\rangle$. The state (\ref{state3}) can now
be written as
\begin{eqnarray}
|\Psi(x)\rangle &=& \tfrac{\alpha}{2}[\psi_1(x) + \psi_2(x)]|q_+\rangle\nonumber\\
&& + \tfrac{\alpha}{2}e^{\frac{i}{\hbar}p_0x}[\psi_1(x) + \psi_2(x)]|q_-\rangle\nonumber\\
 && + \tfrac{\beta}{\sqrt{2}}[\psi_1(x) + \psi_2(x)] |q_3\rangle \nonumber\\
 &=& \Psi_0(x)[\tfrac{\alpha}{\sqrt{2}}|q_+\rangle+ \beta|q_3\rangle] 
+ e^{\frac{i}{\hbar}p_0x}\Psi_0(x) \tfrac{\alpha}{\sqrt{2}}|q_-\rangle\nonumber\\
\label{state4}
\end{eqnarray}
Given the fact that $|q_+\rangle,|q_-\rangle,|q_3\rangle$ form an orthonormal
set, equation (\ref{state4}) implies that the quanton passes undisturbed
whenever the which-way detector state is $|q_+\rangle$ or $|q_3\rangle$. But
when the which-way detector state is $|q_-\rangle$, the quanton \emph{appears
to} experience a momentum kick of magnitude $p_0=h/2d$. Let us see how the state
(\ref{state4}) looks in the momentum basis. It is straightforward to write
\begin{eqnarray}
\Phi(p) = \Phi_0(p)[\tfrac{\alpha}{\sqrt{2}}|q_+\rangle+ \beta|q_3\rangle] 
+ \Phi_0(p-p_0) \tfrac{\alpha}{\sqrt{2}}|q_-\rangle, \nonumber\\
\label{phip}
\end{eqnarray}
where $\Phi_0(p)$ is the momentum representation of the undisturbed quanton
state $\Psi_0(x)$, given by (\ref{ent0}). This form clearly brings out the
apparent momentum kick of magnitude $p_0$. Looking at it, one realizes
that the mistake Storey et. al. \cite{storey} made was in forcing the 
which-way detector to have position degree of freedom, and then
misinterpreting the momentum shift in the quanton as a \emph{momentum transfer}
from the path detector to the quanton. The apparent moment shift is just
an artifact of considering the which-way detector in a different basis.
A similar mistake led others \cite{kick-wiseman1} to conclude that it is
not possible to demonstrate momentum kicks in the proposed experiment of
Scully, Englert and Walther \cite{SEW}, because the microwave cavity
which-way detectors, in their setup, did not involve any position degrees
of freedom. We have demonstrated here that the concept of momentum kicks
is very general and works for any kind of which-way detector.

For imperfect which-way detection, Storey et. al. had derived a relation between
the interference visibility and the \emph{maximum transferred momentum} 
$p_m$ \cite{storey}
\begin{equation}
 \frac{p_m d}{\hbar} \ge 1 - \mathcal{V}.
\end{equation}
This gives the impression that the magnitude of momentum kicks will become smaller
if the which-way detection is imperfect. However, the preceding analysis
shows that the magnitude of the momentum kicks is $h/2d$, and does not
depend on the how good the which-way detection is. So what \emph{does}
depend on the efficiency of which-way detection? Equation (\ref{phip})
says that the probably of the quanton getting a momentum kick is
$\alpha^2/2$. Remembering the value of $\alpha$ from (\ref{betagamma}), we
can write the probability of a quanton getting a kick, or the fraction
of quantons (from a stream) which receive a momentum kick, as
\begin{equation}
 F_k = \tfrac{1}{2}(1 - |\langle d_1|d_2\rangle|).
\end{equation}
For perfect which-way detection $F_k = \tfrac{1}{2}$ which agrees with the
earlier result \cite{tq}. If the which-way detection becomes more imperfect,
the quanton receives apparent momentum kicks of the same magnitude, but
less often. The probability of the quanton receiving momentum-kicks is also 
related to the visibility of interference in a straightforward manner
\begin{equation}
 F_k = \tfrac{1}{2}(1 - \mathcal{V}).
\end{equation}
So, the interference visibility is not related to the magnitude of the
momentum kicks, but to the probability of the quanton receiving a 
momentum kick.

Let us now consider the case where $\theta$ may not be zero, and 
$\delta = e^{i\theta}\beta$. In this case, instead of (\ref{state3}) one
would write
\begin{eqnarray}
|\Psi(x)\rangle &=& \tfrac{\alpha}{2}[\psi_1(x) + \psi_2(x)]|q_+\rangle\nonumber\\
&& + \tfrac{\alpha}{2}[\psi_1(x) - \psi_2(x)]|q_-\rangle\nonumber\\
 && + \tfrac{\beta}{\sqrt{2}}[\psi_1(x) + e^{i\theta}\psi_2(x)] |q_3\rangle.
\label{state5}
\end{eqnarray}
The question one may then ask is, can the factor $e^{i\theta}$ be also
interpreted as a momentum kick, and does it affect the visibility of
interference? It is true that the last term may also be written as
\begin{eqnarray}
|\Psi(x)\rangle &=& \tfrac{\alpha}{2}[\psi_1(x) + \psi_2(x)]|q_+\rangle\nonumber\\
&& + \tfrac{\alpha}{2}[\psi_1(x) - \psi_2(x)]|q_-\rangle\nonumber\\
 && + \tfrac{\beta}{\sqrt{2}}e^{\frac{i}{\hbar}p_e x} [\psi_1(x) + \psi_2(x)] |q_3\rangle,
\label{state6}
\end{eqnarray}
where $p_e = \theta\hbar/d$ may be interpreted as the magnitude of a
momentum kick. However, it is easy to see that it will not affect the
interference visibility. The two interference patterns arising from the
first two terms in (\ref{state6}) cancel each other out exactly, and
the effect of $e^{\frac{i}{\hbar}p_e x}$ on the third interference will
be to just shift it by a fixed amount. While it is clear that the presence
of $e^{\frac{i}{\hbar}p_e x}$  does not affect the interference visibility,
it also highlights the rather artificial nature of the concept of momentum
kicks. It would probably be better to look at it as a phase
difference between the two paths \cite{luis,unni}.

Lastly we probe a situation where the detector states $|q_{\pm}\rangle$
may form any general basis, which is unbiased with respect to
$|q_1\rangle, |q_2\rangle$. For example, let us assume them to have the form
$|q_{\pm}\rangle = \tfrac{1}{\sqrt{2}}(|q_1\rangle \pm e^{i\theta} |q_2\rangle)$,
where $\theta$ is an arbitrary angle.
Using this basis, the state in (\ref{state2}) can be written as
\begin{eqnarray}
|\Psi(x)\rangle &=& \tfrac{\alpha}{2}[\psi_1(x) + e^{-i\theta}\psi_2(x)]|q_+\rangle\nonumber\\
&& + \tfrac{\alpha}{2}[\psi_1(x) - e^{-i\theta}\psi_2(x)]|q_-\rangle\nonumber\\
 && + \tfrac{\beta}{\sqrt{2}}[\psi_1(x) + \psi_2(x)] |q_3\rangle.
\label{nstate3}
\end{eqnarray}
Since the two quanton states correlated with $|q_{\pm}\rangle$ differ from
each other only in a phase difference of $\pi$ between the two paths,
it is straightforward to see that the state 
$\tfrac{1}{\sqrt{2}}[\psi_1(x) - e^{-i\theta}\psi_2(x)]$ is the same as 
$e^{\frac{i}{\hbar}p_0x}\tfrac{1}{\sqrt{2}}[\psi_1(x) + e^{-i\theta}\psi_2(x)]$,
where $p_0=h/2d$ is a momentum kick. However, if one insists on calculating
the momentum kicks with respect to the original state without any path
detector, $\tfrac{1}{\sqrt{2}}[\psi_1(x) + \psi_2(x)]$, the momentum kicks
come out to be $\hbar\theta/d$ and $\hbar\theta/d + h/2d$. So, there appears
to be an ambiguity in the magnitude of the momentum kicks. However, notice
that the momentum kick difference between the two states is still $h/2d$.
Thus to avoid ambiguity, the momentum kicks should only be interpreted
as \emph{relative} kicks between the two states of the quanton. This
aspect is easily understood by recalling that in the phenomenon of quantum
erasure, the recovered interference may be shifted from the center, depending
on which basis of the which-way detector one is using \cite{nochoice}.

\section{Conclusion}

With the aim of clarifying the long-standing controversial issue
of momentum kicks in which-way detection,
we have theoretically analyzed a two-slit interference experiment with
imperfect which-way detection, using the concept of UQSD. We have shown
that the partial loss of
interference may be interpreted as arising due to random momentum kicks
which the quanton \emph{appears to} experience. Interestingly, the
\emph{relative} magnitude
of the momentum kicks always remains the same, $h/2d$, irrespective of
the efficiency of which-way detection. Earlier arguments seemed to suggest
otherwise \cite{storey}. However, as the which-way detection
becomes more imperfect, the momentum kicks become less frequent. In the
case of perfect which-way detection, the quanton experiences momentum
kicks fifty percent of the time.

While the analysis presented here uses UQSD only as a tool to clarify the
concept of momentum kicks, all this can be easily experimentally tested,
as unambiguous path discrimination in two-path interference has already 
been experimentally demonstrated \cite{neves,len}.

The lesson learnt here is that the roots of complementarity lie in the
ubiquitous entanglement between the quanton and the which-way detector.
When the which-way detector is looked at in one particular basis, 
the loss of interference can be interpreted as arising from the quantum
correlation between the two paths and two states of the which-way detector.
If the which-way detector is looked at in another mutually unbiased basis,
the loss of interference can be interpreted as arising from the quanton
experiencing random momentum kicks of magnitude $h/2d$. However, it should
be emphasized that these are only apparent momentum kicks arising out of
a phase difference between the two paths. There is no real momentum transfer
to the quanton from anywhere. We hope this finally puts the controversy
surrounding the issue of momentum kicks at rest.

\begin{acknowledgements}
Neha Pathania acknowledges financial support from the Department of Science
and Technology, through the Inspire Fellowship (registration number IF180414).
\end{acknowledgements}
\onecolumngrid
\vskip 5mm

\centerline{\rule{5cm}{0.4pt}}
\vskip 5mm
\twocolumngrid

\onecolumngrid

\begin{thebibliography}{0}
\bibitem{bohr} N. Bohr, ``The quantum postulate and the recent development of
atomic theory," 
\href{ https://doi.org/10.1038/121580a0}{\emph{Nature} (London) 
 {\bf 121}, 580-591 (1928)}.

\bibitem{recoil} N. Bohr, in \emph{ Albert Einstein: Philosopher-Scientist}
(ed. Schilpp, P. A.) 200-241 (Library of Living Philosophers, Evanston, 1949);
reprinted in Quantum Theory and Measurement (eds J.A. Wheeler, W.H. Zurek,)
9-49 (Princeton Univ. Press, 1983).

\bibitem{recoillui} X-J Liu, Q. Miao, F. Gel'mukhanov, M. Patanen, O. Travnikova, C. Nicolas, H. Agren, K. Ueda, C. Miron
``Einstein–Bohr recoiling double-slit gedanken experiment performed at the molecular level,"
\href{https://doi.org/10.1038/nphoton.2014.289}
{\emph{Nature Photonics} \textbf{9}, 120-125 (2015).}

\bibitem{recoildorner} L.Ph.H. Schmidt, J. Lower, T. Jahnke, S. Sch\"oßler, M.S. Sch\"offler, A. Menssen, C. L\'ev\^eque,
N. Sisourat, R. Ta\"ieb, H. Schmidt-Böcking, and R. D\"orner,
``Momentum Transfer to a Free Floating Double Slit: Realization of a Thought Experiment
from the Einstein-Bohr Debates,"
\href{https://doi.org/10.1103/PhysRevLett.111.103201}
{\emph{ Phys. Rev. Lett.} \textbf{ 111}, 103201 (2013).}

\bibitem{utter} R.S. Utter, J.M. Feagin,
``Trapped-ion realization of Einstein's recoiling-slit experiment",
\href{https://doi.org/10.1103/PhysRevA.75.062105}
{\emph{ Phys. Rev. A} \textbf{ 75}, 062105 (2007).}

\bibitem{SEW} M.O. Scully, B.G. Englert, H. Walther,
``Quantum optical tests of complementarity,"
\href{https://doi.org/10.1038/351111a0}
{\emph{ Nature} \textbf{ 351}, 111-116 (1991).}

\bibitem{storey} E.P. Storey, S.M. Tan, M.J. Collett, D.F. Walls, ``Path detection and the uncertainty principle,"
\href{https://doi.org/10.1038/367626a0}
{\emph{ Nature} \textbf{ 367}, 626-628, (1994).}

\bibitem{kick-englert} B.-G. Englert, M.O. Scully, H. Walther, ``Complementarity
and uncertainty,"
\href{https://doi.org/10.1038/375367b0}
{\emph{ Nature} \textbf{ 375}, 367 (1995).}

\bibitem{kick-storey} E.P. Storey, S.M. Tan, M.J. Collett, D.F. Walls, ``Complementarity and uncertainty,"
\href{https://doi.org/10.1038/375368a0}
{\emph{ Nature} \textbf{ 375}, 368 (1995).}

\bibitem{kick-wiseman1} H. Wiseman, F. Harrison, ``Uncertainty over complementarity?"
\href{https://doi.org/10.1038/377584a0}
{\emph{ Nature} \textbf{ 377}, 584 (1995).}

\bibitem{kick-wiseman2} H.M. Wiseman, F.E. Harrison, M.J. Collett, S.M. Tan, D.F. Walls, R.B. Killip, ``Nonlocal momentum transfer in welcher Weg measurements," 
\href{https://doi.org/10.1103/PhysRevA.56.55}
{\emph{ Phys. Rev. A} \textbf{ 56}, 55 (1997).}

\bibitem{kick-wiseman3} H.M. Wiseman, ``Bohmian analysis of momentum transfer in welcher Weg measurements,"
\href{https://doi.org/10.1103/PhysRevA.58.1740}
{\emph{ Phys. Rev. A} \textbf{ 58}  1740 (1998).}

\bibitem{wiseman2}  H. Wiseman, ``Directly observing momentum transfer in
twin-slit which-way experiments"
\href{https://doi.org/10.1016/S0375-9601(03)00504-8}
{\emph{ Phys. Lett. A} \textbf{ 311}, 285 (2003).}

\bibitem{englert} B-G. Englert, ``Fringe visibility and which-way information:
an inequality",
\href{https://doi.org/10.1103/PhysRevLett.77.2154} {\emph{Phys. Rev. Lett.} {\bf 77}, 2154 (1996).}

\bibitem{tqrv} T. Qureshi, R. Vathsan, ``Einstein's recoiling slit experiment, complementarity and uncertainty,"
\href{https://doi.org/10.12743/quanta.v2i1.11}
{\emph{Quanta} \textbf{2}, 58-65 (2013).}

\bibitem{durr} S. Dürr, G. Rempe, ``Can wave–particle duality be based on the uncertainty relation?"
\href{https://doi.org/10.1119/1.1285869}
{\emph{ Am. J. Phys.} \textbf{ 58}, 1021-1024 (2000).}

\bibitem{mir}  R. Mir, J.S. Lundeen, M.W. Mitchell, A.M. Steinberg, J.L. Garretson, H.M. Wiseman, ``A double-slit 'which-way' experiment on the complementarity–uncertainty debate,"
\href{https://doi.org/10.1088/1367-2630/9/8/287}
{\emph{ New J. Phys.} \textbf{ 9}, 287 (2007).}

\bibitem{xiao}  Y. Xiao, H.M. Wiseman, J-S. Xu, Y. Kedem, C-F. Li, G-C. Guo, ``Observing momentum disturbance in double-slit 'which-way' measurements"
\href{https://doi.org/10.1126/sciadv.aav9547}
{\emph{ Sci. Adv.} \textbf{ 5}, eaav9547 (2019).}

\bibitem{tq} T. Qureshi, ``Which-way measurement and momentum kicks,"
\href{https://doi.org/10.1209/0295-5075/123/30007}
{\emph{ EPL} \textbf{ 123}, 30007 (2018).}

\bibitem{awpd} K. Menon, T. Qureshi, ``Wave-particle duality in asymmetric beam interference,"
\href{https://doi.org/10.1103/PhysRevA.98.022130}
{\emph{ Phys. Rev. A} \textbf{ 82}, 022130 (2018).}

\bibitem{uqsd}  I.D. Ivanovic, ``How to differentiate between non-orthogonal
states",
\href{https://doi.org/10.1016/0375-9601(87)90222-2}
{\emph{ Phys. Lett. A} \textbf{ 123}, 257 (1987).}

\bibitem{dieks}  D. Dieks, ``Overlap and distinguishability of quantum states,"
\href{https://doi.org/10.1016/0375-9601(88)90840-7}
{\emph{ Phys. Lett. A} \textbf{ 126}, 303 (1988).}

\bibitem{peres}  A. Peres, ``How to differentiate between non-orthogonal
states ,"
\href{https://doi.org/10.1016/0375-9601(88)91034-1}
{\emph{ Phys. Lett. A} \textbf{ 128}, 19 (1988).}

\bibitem{jaeger2} G. Jaeger, A. Shimony, ``Optimal distinction between
two non-orthogonal quantum states,"
\href{https://doi.org/10.1016/0375-9601(94)00919-G}
{\emph{ Phys. Lett. A} \textbf{ 197}, 83 (1995).}

\bibitem{luis} A. Luis, L.L. Sánchez-Soto, ``Complementarity enforced by random classical phase kicks,"
\href{https://doi.org/10.1103/PhysRevLett.81.4031}
{\emph{ Phys. Rev. Lett.} \textbf{ 81}, 4031 (1998).}

\bibitem{unni} C.S. Unnikrishnan, ``Origin of quantum-mechanical
complementarity without momentum back action in atom-interferometry
experiments,"
\href{https://doi.org/10.1103/PhysRevA.62.015601}
{\emph{ Phys. Rev. A} \textbf{ 62}, 015601 (2000).}

\bibitem{nochoice} T. Qureshi, ``The delayed-choice quantum eraser leaves no choice,"
\href{https://doi.org/10.1007/s10773-021-04906-w}
{\emph{Int. J. Theor. Phys.} (2021)}

\bibitem{neves} L. Neves, G. Lima, J. Aguirre, F. A. Torres-Ruiz, C. Saavedra, A. Delgado, ``Control of quantum interference in the quantum eraser,"
\href{https://doi.org/10.1088/1367-2630/11/7/073035}
{\emph{New J. Phys.} \textbf{11}, 073035 (2009).}

\bibitem{len} Y. L. Len, J. Dai, B-G Englert, L.A. Krivitsky, ``Unambiguous path discrimination in a two-path interferometer,"
\href{https://doi.org/10.1103/PhysRevA.98.022110}
{\emph{ Phys. Rev. A} \textbf{98}, 022110 (2018).}

\end{thebibliography}
\end{document}